\documentclass[twocolumn,showpacs,preprintnumbers,amsmath,amssymb]{revtex4} 

\usepackage{graphicx}
\usepackage{dcolumn}
\usepackage{bm}

\begin{document} 


\title{Effective Temperatures in Driven Systems:  Static 
vs.~Time-Dependent Relations} 
\author{Corey S. O'Hern} 
\affiliation{ Department of Mechanical Engineering, 
Yale University, New Haven, CT 06520-8284.} 
\author{Andrea J. Liu} 
\affiliation{Department of Chemistry and Biochemistry, UCLA, Los 
Angeles, CA 90095-1569} 
\author{Sidney R. Nagel} 
\affiliation{James Franck Institute, The University of Chicago, 
Chicago, IL 60637} 
\date{\today} 

\begin{abstract} 
Using simulations of glassy systems under steady-state shear, we 
compare effective temperatures obtained from static linear response 
with those from time-dependent fluctuation-dissipation relations. 
Although these two definitions are not expected to agree, we show that 
they yield the same answer over two and a half decades 
of effective temperature.  This suggests that a more complete 
conceptual framework is necessary for effective temperatures in 
steady-state driven systems. 
\end{abstract} 
\maketitle 

Temperature is one of the fundamental variables in an equilibrium
system that determines not only the system's average properties, such
as pressure or density, but also fluctuations around those
averages. Temperature also relates, via linear response,
fluctuations in a thermodynamic quantity to that quantity's response
to a small perturbation in its conjugate variable. When a system is
far out of equilibrium, temperature is no longer well-defined.
Nevertheless, in many cases there will still be fluctuations although
they are not thermal in origin.  An example of this is a steady-state
driven system such as a sheared material where shear introduces
fluctuations that are not described by a thermal bath temperature.
Can one define an appropriate ``effective temperature'' to
characterize these fluctuations?  

\begin{figure} 
\scalebox{0.35}{\includegraphics{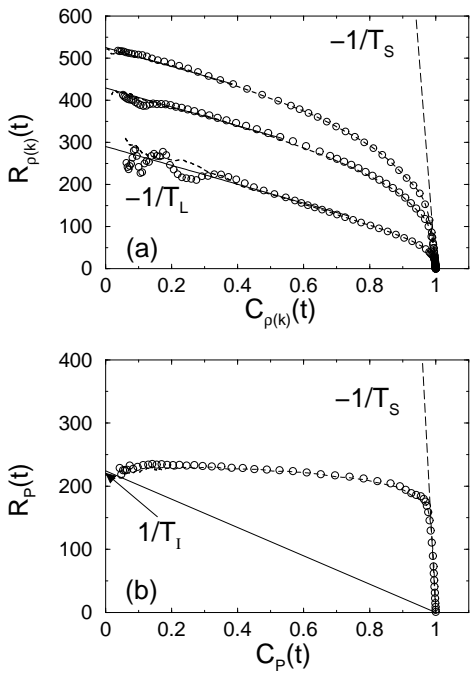}}%
\vspace{-0.15in} 
\caption{(a) $R_{\rho(k)}(t)$ vs.~correlation
$C_{\rho( k)}(t)$ in a 2d system with repulsive harmonic interactions
at a bath temperature $T_{KE} =10^{-4}$, shear rate ${\dot
\gamma}=0.01$, and packing fraction $\phi=0.90$ for $N=256$ particles.
The wavevector ${\vec k}$ lies in the shear gradient direction;
$k=4.5$, $9$, and $11$ are shown with $k$ decreasing from top to
bottom.  The solid lines are guides to the eye and have slopes equal
to $-1/T_L$.  (b) $R_P(t)$ vs.~$C_P(t)$ for pressure in the same
system as (a). The solid line is a guide to the eye. It has slope
$-1/T_L$ and intercept $1/T_L$ where $T_L$ is obtained from (a). Note
that $T_I$ from (b) is equal to $T_L$ from (a). The results are in the
linear response regime, as shown by the open circles and dashed lines
in each of the figures, which are for two magnitudes of the perturbing
field that differ by a factor of $5$.}
\label{figure1} 
\vspace{-0.2in} 
\end{figure} 

For the idea of an effective temperature to be useful, a clear
prescription for defining it should exist and this prescription should
apply generally in different contexts.  Various groups have defined
and measured different effective temperatures in systems far from
equilibrium \cite{jou,barrat,invariant,cugliandolo2}.  We will
show here that two prescriptions based on linear response that have
seemed to be fundamentally incompatible do, surprisingly, give the
same value for the effective temperature.  For this to occur, neither
prescription can work infallibly in all cases.  Thus, our results pose
the conceptual question: under which conditions should either linear
response prescription be applied?

Linear response provides a large number of possible definitions for
the effective temperature, each based on a different pair of conjugate
variables, which all reduce to the true temperature in thermal
equilibrium.  Effective temperatures based on these different
relations have been used to model simulations of particulate systems
driven out of equilibrium by steady-state
shear~\cite{berthier,makse,ono}.  There are two ways in which such
relations have been implemented with significant success.  One way
uses static linear response which relates equal-time fluctuations to
the response at {\it infinite-times} to yield an effective temperature
$T_{I}$.  The other way, argued to be more
fundamental\cite{kurchan,berthier2}, measures the autocorrelation
function and relates it to the response {\it as a function of time}.
In equilibrium there is a strict proportionality between correlation
and response at all times and therefore a single well-defined
temperature, but in driven systems the situation is more subtle.  The
conceptual picture behind this is that there can be two widely
separated time scales in the presence of shear.  Degrees of freedom
that decay on a short time scale are characterized by one effective
temperature, $T_{S}$, while those that take a longer time to decay are
characterized by a different (higher) value, $T_{L}$.  It has been
predicted that $T_{S}$ is the bath temperature because fast degrees of
freedom decay before shear has any effect, while $T_{L}$, which can
only be obtained from the long-time behavior of correlation and
response, corresponds to a well-defined effective temperature
characterizing structural rearrangements driven by
shear~\cite{kurchan}.

As will be made clear below, an effective temperature based on 
infinite time, or static linear response, $T_{I}$, should generally 
not agree with $T_{L}$ obtained from time-dependent linear response in 
non-equilibrium systems.  Indeed, for the same pair of conjugate 
variables we always find in our simulations that $T_{I} \ne T_{L}$. 
However, for a wide variety of simulations we find that $T_{I}$ for 
one conjugate pair can be equal to $T_{L}$ for a different pair. 
Therefore, it appears that under different conditions both definitions 
of effective temperature must be equally valid.  The conceptual 
framework~\cite{cugliandolo,cugliandolo2} that has been used to argue   
for the 
validity of $T_{L}$ would imply that $T_{I}$ should never be valid, in 
contradiction to our findings.  That scenario must therefore be 
incomplete. 

In order to demonstrate these results, we have performed numerical
simulations of systems undergoing linear shear flow in both two and
three spatial dimensions ($2d$ and $3d$) under conditions of fixed volume,
fixed number of particles and fixed shear rate.  The systems are
composed of $50$-$50$ bidisperse mixtures with diameter ratio $1.4$,
which prevents crystallization and segregation.  The system is
enclosed in a cubic simulation cell with Lees-Edwards periodic
boundary conditions to impose shear in the $x$-direction and a shear
gradient in the $y$-direction. Particles interact via one of the
following pairwise, finite-range, purely repulsive potentials:
\begin{eqnarray} 
\label{potentialdef} 
V^{hs}(r_{ij}) & \equiv & \frac{\epsilon}{2} 
\left(1-r_{ij}/\sigma_{ij}\right)^2 \nonumber \\ 
V^{H}(r_{ij}) & \equiv & \frac{2\epsilon}{5} 
\left(1-r_{ij}/\sigma_{ij}\right)^{5/2} \\ 
V^{RLJ}(r_{ij}) & \equiv & \frac{\epsilon}{72}\left[ 
({\sigma_{ij}/r_{ij}})^{12} - 2({\sigma_{ij}/r_{ij}})^6 + 1\right],   
\nonumber 
\end{eqnarray} 
where $\epsilon$ is the characteristic energy scale of the
interaction, $\sigma_{ij}$ is the average diameter of particles $i$
and $j$, and $r_{ij}$ is their separation.  All potentials (harmonic
spring, Hertzian nonlinear spring, and repulsive Lennard-Jones) are
zero when $r_{ij} \ge \sigma_{ij}$.  Our results have been obtained
with packing fractions ranging from $\phi=[0.70,0.84]$ in $3d$ and
$\phi=[0.85,1.20]$ in $2d$, which are all above random
close-packing\cite{longJ}.  We varied the number of particles in the
range $N=[256,1024]$ and found no appreciable finite-size effects for
the results reported here.  The units of length, energy, and time are
$\sigma$, $\epsilon$, and $\sigma \sqrt{m/\epsilon}$, respectively
where $m$ is the particle mass and $\sigma$ is the small-particle
diameter.

We have studied both thermal and athermal (or dissipative) systems to 
show that our results are {\it not} specific to any particular 
dynamics.  Thermal systems under shear can be described 
by the Sllod equations of motion for the position ${\vec r}_i$ and 
velocity fluctuation ${\vec v}_i$ of each particle around the average 
linear velocity profile \cite{evans}: 
\begin{equation} 
\label{sllod} 
\frac{d{\vec r}_i}{dt} = {\vec v}_i + 
{\dot \gamma} y_i {\hat x} , 
\frac{d{\vec v}_i}{dt} = 
{\vec F}^r_i/m - {\dot \gamma} v_{yi} {\hat x} - \alpha {\vec v}_i, 
\end{equation} 
where ${\vec F}^r_i=-\sum_j dV(r_{ij})/dr_{ij} {\hat r}_{ij}$ is the 
repulsive force on particle $i$ due to neighboring particles 
$j$, ${\dot \gamma}$ is the shear rate, and $\alpha$ is chosen to fix 
the kinetic energy per degree of freedom, $T_{KE}$, associated with 
velocity fluctuations.  We always set $T_{KE}$ to be 
below the glass transition temperature of the unsheared 
system. 

Athermal dissipative systems can be described by\cite{luding}: 
\begin{equation} 
\label{dissipative} 
m \frac{d^2{\vec r}_i}{dt^2} = {\vec F}^r_i - b \sum_j ({\vec v}^t_{i} 
- {\vec v}^t_{j}), 
\end{equation} 
where ${\vec v}^t_{i}$ is the total velocity (including shear) of 
particle $i$, $b>0$ is the damping coefficient, and the sum over $j$ 
only includes particles that overlap particle $i$. At finite shear 
rate, these systems reach a steady-state where the power put in by the 
shear flow balances the power dissipated.  In this study, we focused 
on underdamped dissipative dynamics and therefore fixed the 
dimensionless damping coefficient $b^* = b \sigma/\sqrt{\epsilon m} 
\ll 1$. 

\begin{figure} 
\scalebox{0.35}{\includegraphics{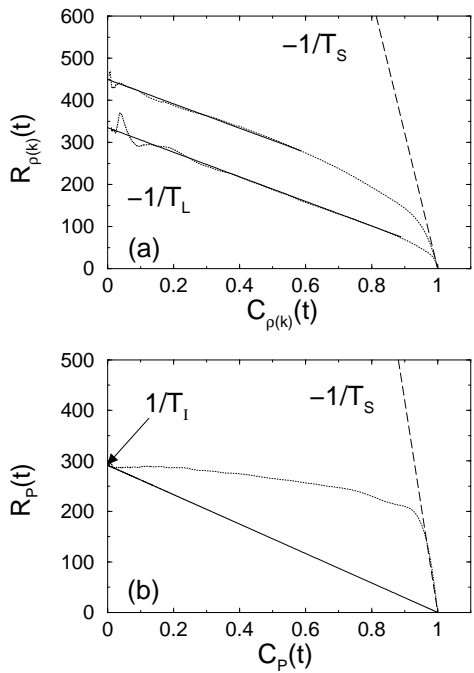}}%
\vspace{-0.15in} 
\caption{(a) $R_{\rho(k)}(t)$ plotted versus correlation $C_{\rho(
k)}(t)$ in a $2d$ sheared, {\it athermal} system with repulsive harmonic
interactions at $\phi=0.90$ and ${\dot \gamma}=0.01$.  Again, ${\vec
k}$ lies in the shear gradient direction and $k=4.5$ and $9$ are shown
with $k$ decreasing from top to bottom.  The solid lines have slope
equal to $-1/T_L$. (b) $R_P(t)$ plotted versus the autocorrelation
function $C_P(t)$ for pressure $P$ in the same system as (a).  The
straight line has slope $-1/T_L$ and intercept $1/T_L$ where $T_L$ is
obtained from (a).}
\label{figure2} 
\vspace{-0.2in} 
\end{figure} 

We now describe how to calculate the effective temperature from a set 
of conjugate variables and then show explicitly the incompatibility of 
the static and the time-dependent definitions for the same conjugate 
pair of variables.  Since we are concerned with systems in 
steady-state shear, we assume a final steady state in which averaged 
quantities become time independent.  Consider an observable, $A(t)$, 
that fluctuates in time $t$, such as the number 
density or the total pressure of the system.  Then one can define the 
autocorrelation function: 
\begin{equation} 
\widetilde C_{A}(t)= \langle A(t) A(0) \rangle - 
\langle A \rangle^{2} 
\label{corrdef} 
\end{equation} 
where $\langle A \rangle$ represents an average of $A$ over 
time and configurations.  If $B$ is the thermodynamic field 
conjugate to $A$, then one can also define an integrated response 
function which measures the response to a small constant 
perturbation, $\delta B$, applied from time $t=0$ onward: 
\begin{equation} 
\widetilde R_{A}(t)= \frac{\langle A(t) \rangle - \langle 
A(0) \rangle}{\delta B} 
\label{respdef} 
\end{equation} 
If we now introduce the rescaled variables 
\begin{equation} 
\label{eq_R} 
R_A(t)=\frac{\widetilde R_A(t)}{\widetilde C_A(t=0)}~{\rm and}~ 
C_A(t)=\frac{\widetilde C_A(t)}{\widetilde C_A(t=0)}, 
\end{equation} 
the fluctuation-dissipation relation states that 
\begin{equation} 
R_A(t)= \frac{1}{T} \left( 1-C_A(t) \right). 
\label{fdt} 
\end{equation} 
This implies that if $R_A$ is plotted parametrically against $C_A$, 
the result should be a straight line with slope $-1/T$ for an 
equilibrium system.  Moreover, the infinite-time limit, which is the 
intercept of such a plot on the $R$-axis, has the value $1/T$. 
According to Eqs.~\ref{eq_R} and \ref{fdt}, the intercept satisfies 
\begin{equation} 
\label{eq_intercept} 
\widetilde R_A(t=\infty) = \widetilde C_A(t=0)/T 
\end{equation} 
Thus, the intercept, which defines a temperature $T_{I}$, corresponds 
to the static linear response relation whereby the infinite-time 
response, $\widetilde R_A(t=\infty)$ (Eq.~\ref{respdef}), is related 
to the equal-time correlation function, $\widetilde C_{A}(t=0)$ 
(Eq.~\ref{corrdef}). 

As indicated earlier, Kurchan~\cite{kurchan} has predicted that for 
driven systems in steady state such a 
parametric plot has two regimes~\cite{kurchan,berthier}.  Berthier and 
Barrat have conducted simulations of sheared Lennard Jones glasses and   
shown 
that at short times ($C_A$ close to one and $R_A$ close to zero), the 
slope of the line, $-1/T_S$, defines a temperature characterizing the 
fast modes in the system and corresponds to the bath temperature,   
$T_{KE}$.  At long times 
the slope, $-1/T_L$, is a good measure of the 
effective temperature produced by the shear for the slow modes.  We 
show a similar plot for a $2d$ sheared thermal system in 
Fig.~\ref{figure1}(a) using as the variable the Fourier component of 
the number density of the large particles at various values of the 
wavevector, ${\vec k}$: 
\begin{equation} 
\label{eq_rho_k} 
\rho({\vec k},t) = \sum_{i=1}^{N/2} e^{i {\vec k} \cdot {\vec r}_i(t)}. 
\end{equation} 
For this variable, we calculate the incoherent part of the scattering   
function 
for the large particles, 
$C_{\rho( k)}(t)$ \cite{berthier}. 
In Fig.~\ref{figure1}(a), there is a well-defined slope at short times   
and a smaller slope at 
long times (i.e., at smaller values of $C_A(t)$).  It is quite obvious 
that the long-time slopes are all the same so that there is a common 
effective temperature that describes the fluctuation and response for 
all of these variables.  These results are completely consistent with 
the results found by Berthier and Barrat \cite{berthier} for a 
three-dimensional system.  Note that in all of these cases, the value 
at which each of these curves intercepts the $R_{\rho({k})}$ axis 
cannot have the value $1/T_L$.  This could only be the case if $T_S = 
T_L$ (as in equilibrium), or if the regime corresponding to 
$T_S$ shrinks to zero~\cite{mayer}.  If the curve is not a straight 
line, then $T_{I}$ (determined from the 
intercept) must be different from {\it either} $T_S$ or $T_L$. 

In Fig.~\ref{figure1}(b), we show the parametric plot for the 
identical system as in Fig.~\ref{figure1}(a) but for a different 
variable, namely the total pressure, $P$.  We 
calculate the pressure $P=P_{\alpha \alpha}/d$ in $d$ spatial 
dimensions, using the following expression for the pressure tensor 
\begin{equation} 
\label{pressure} 
L^d P_{\alpha \beta} = \sum_{i=1}^N m v_{\alpha i} v_{\beta i} + 
\sum_{i=1}^{N-1} \sum_{j=i+1}^N r_{\alpha ij} F^r_{\beta ij}, 
\end{equation} 
where $\alpha$,$\beta=x$,$y$, or $z$ and $L$ is the edge length of the
simulation box.  The shape of this curve is very different from those
shown in Fig.~\ref{figure1}(a).  In this case, the response rises
rapidly at short times and then turns over and becomes horizontal at
long times~\cite{flat,cugliandolo2}. The striking result is that,
although this curve is manifestly different from those shown earlier,
it has an intercept temperature, $T_{I}$, with the same value,
$T_{I}=T_L$, obtained from the late-time slopes of the curves in
Fig.~\ref{figure1}(a).

\begin{figure} 
\scalebox{0.4}{\includegraphics{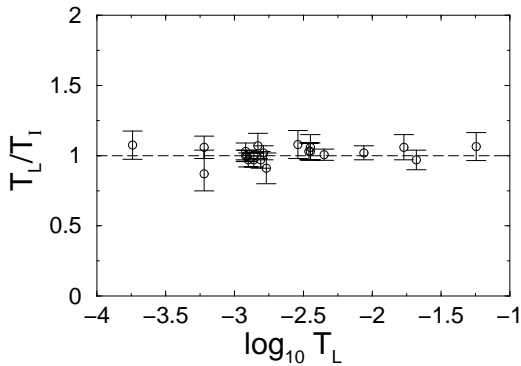}}%
\vspace{-0.15in} 
\caption{The ratio $T_L/T_I$ is plotted vs. $T_L$, where $T_I$ is
obtained from the $R$-intercept of integrated response (R)
vs. correlation (C) for pressure and $T_L$ is obtained from the
long-time slope of R vs. C for $\rho(k)$. The largest error comes
from estimating the long-time slope of R vs. C for $\rho(k)$.}
\label{figure3} 
\vspace{-0.2in} 
\end{figure}

We stress here that this is not a coincidence.  In Fig.~\ref{figure2}
we show for an athermal system the response versus correlation plots
for the same sets of variables as shown for thermal systems in
Fig.~\ref{figure1}.  Again we see that $T_{I}$ obtained from $P$ has
the same value as $T_{L}$ obtained from $\rho({k})$.  To indicate the
full extent of agreement we plot in Fig.~\ref{figure3} the ratio
$T_{L}/T_I$ versus $\log_{10} T_L$ for all systems studied.  This figure
shows that within error $T_L=T_I$ over two and half decades of
effective temperature. These data are collected from thermal and
athermal simulations at different values of the shear rate and density
in $2d$ and $3d$ for systems with different particle interactions. For
the thermal simulations, we also varied the bath temperature.  

We have shown that both static linear response and time-dependent
fluctuation-dissipation relations can yield consistent values of
effective temperature. However, it is unclear when one should use
static relations and when one should use time-dependent ones.
Pressure is not the only observable for which the static relation is
appropriate.  Previously, we showed that static relations yield a
consistent effective temperature for shear stress and potential
energy, as well \cite{ono}.  It is also not true that one should
invariably use the static relation for quantities involving pressure.
We have calculated response vs. correlation for pressure at different
values of ${\vec k}$ in $2d$ sheared systems.  At large $k$, $T_L$ 
yields the correct effective temperature whereas at
$k=0$ $T_I$ does.

It is not true that the effective temperature for all zero-wavevector
observables should be given by static linear response.  While static
linear response appears appropriate for $k=0$ pressure, energy and
shear stress, we find other examples for which this does not hold,
including $P_{zz}$ \cite{berthier} and the deviatoric pressure
$P_{dev} = (2P_{11} - P_{22} - P_{33})/3$ where $1,2,3=x,y,z$
\cite{parisi}.  We find that the response-correlation curves for each
of these $k=0$ variables are not flat as in Fig.~\ref{figure1}(b), but
have nonzero long-time slopes.  However, the corresponding effective
temperatures can be off by factors of $5-10$ from those
shown in Fig.~\ref{figure1}(a) even when other effective temperatures
agree.

There are regimes where the idea of an effective temperature is valid
and regimes where it breaks down~\cite{uslater}.  For example, at high
densities (e.g. $\phi \approx 1.1$, typical of liquids) and high bath
temperatures ($T>0.1 T_g$) we find that all the different effective
temperatures are the same.  However, these begin to deviate as the
bath temperature is lowered.  Previous studies
\cite{berthier2,kurchan} have suggested that the concept of effective
temperature should be valid when there is a clear separation between
the short-time (bath temperature) regime and the long-time,
shear-rate-dependent regime.  This criterion cannot be sufficient
because as the bath temperature is lowered, the separation between
these two time scales does not decrease.

We have shown that $T_I$ for the zero-wavevector pressure is equal to 
$T_L$ for $\rho(k)$ over a range of two and a half decades in 
effective temperature.  This remarkable result suggests that static 
as well as time-dependent linear response relations can be used to 
define a consistent effective temperature, in contradiction to 
expectations based on spin models subjected to non-conserved 
fields\cite{berthier2,kurchan}.  These results also leave us with a 
puzzle: when should one use static linear response and when should one 
use a time-dependent relation?  For a given pair of conjugate 
variables, there is no obvious criterion for which of these two kinds 
of relations should be used. 

We thank L. Cugliandolo, D. Durian, J. Kurchan, M. Robbins, P.
Sollich, and N. Xu for useful comments.  Grant support from
NSF-DMR-0087349 (CSO,AJL), DE-FG02-03ER46087 (AJL), NSF-DMR-0089081
(CSO,SRN) and DE-FG02-03ER46088 (SRN) is gratefully acknowledged.

\end{document}